\documentclass[fleqn,twoside]{article}
\usepackage{espcrc2}
\usepackage{graphicx}

\title{A Double Parton Scattering Background to Associate $WH$ and $ZH$ Production at the LHC}

\author{M. Y. Hussein\address{Department of Physics, College of Science,
  University of Bahrain, \\ 
  P.O. Box 32038, Kingdom of Bahrain}}

\begin{document}

\begin{abstract}

Higgs boson production in association with $W$ and $Z$ bosons at high
luminosity CERN Large Hadron Collider (LHC,$\sqrt{s}$=14 TeV), is one
of the most promising discovery channel for a SM Higgs particle with a
mass below 135 GeV, where the Higgs decays into $b\bar{b}$ final
states is dominant.  The experimental capability of recognizing the
presence of $b$ quarks in a complex hadronic final state has brought
attention towards the final states with pairs for observing the
production of the Higgs at the LHC. We point out that double parton
scattering processes are going to represent a sizable background to
the process.

\vspace{1pc}
\end{abstract}

\maketitle   % typeset front matter (including abstract)

\section{INTRODUCTION}

The Standard Model beautifully describes all the observed properties
of elementary particles and their interactions. According to this
model, the fundamental particles are either matter particles (the
quarks and leptons) or force particles (the gauge bosons corresponding
to the EM, Weak and Strong forces).

If nature had chosen the most elegant and symmetric mathematical
construct for the SM, all fundamental particles would have zero
mass. Experiments show that this is not the case: all particles have a
small mass.

The Standard Model has no mechanism that would account for any of
these masses, unless we supplement it by adding additional fields, of
a type known as scalar fields.

One of the greatest achievements of twentieth-century theoretical
physics was the discovery of how to incorporate particle masses into
the Standard Model theory without spoiling the symmetry and
mathematical consistency. The most straightforward way to do that is
through what is known as ``Higgs mechanism''~\cite{Hgg64}

An important by-product of this mass generating addition to the
Standard Model is the prediction of the existence of a totally new
type of particle-the as-yet- undiscovered ``Higgs boson''. The
discovery of the Higgs boson would solve the mystery of how particles
get their masses. The spontaneous breakdown of the
weak-electromagnetic gauge group can be accomplished in only one known
way ``Higgs fields''.

The Higgs is elusive and plays a negligible role in low energy
phenomenology because it couples with particles according to their
masses, and so couples very weakly coupled to the lepton and quark
constituents of ordinary matter.

There are a few production mechanisms for the Higgs boson which lead
to detectable cross section at the LHC. Each of them makes use of the
preference of the SM Higgs which couples to the heavy particles,
either massive vector boson or massive quark.

Here we discuss the production of Higgs particle more relevant to an
experimental search ``association production of an intermediate vector
boson $W$ or $Z$ and a Higgs particle''

The $b\bar{b}$ channel is the most favorable Higgs boson decay mode
when the Higgs mass is below the $W^+W^-$ threshold, and since $b$
quark jets are identified with high efficiency final state then
$b\bar{b}$ pairs have a great potential for discovering the Higgs
boson at the LHC, if the Higgs mass is in the range 80 GeV $\leq\ m_H\
\leq$ 150 GeV~\cite{Dai95}.

An effective way to reduce the huge QCD background is to look for $b\bar{b}$
pairs accompanied by isolated lepton resulting from the decay of a $W$
bosons.

At the LHC, Higgs boson production in association with $W$ and $Z$
bosons is the most interest process for detecting the Higgs boson
production through the $b\bar{b}$ decay channel is therefore: \( p+p
\rightarrow W^\pm\, (or\, Z) + X \), with \(W \rightarrow l\nu_l\),
\(Z \rightarrow l\bar{l}\), \(H \rightarrow b\bar{b}\), where
\(l=e,\mu\).

The purpose of the present work is to point out that the same $l$,
$b\bar{b}$ final state can be produced also by double parton
scattering collision processes, which represents therefore a further
background to be taken into account in addition to the other
background processes.

%%%%%%%%%%%%%%%%%%%%%%%%   S e c t i o n   2   %%%%%%%%%%%%%%%%%%%%%%%%

\section{DOUBLE PARTON SCATTERING MECHANISM}

Multiple parton interaction processes, where two different pairs of
partons interact independently at different points in the transverse
space, in the same inelastic hadronic event, become experimentally
important at high energies and with the growing flux of partons.
Double parton scattering has been recently observed by CDF~\cite{Lnd82} .

The parton distribution function depends on the fractional momenta of
all the interacting partons and on their distance in transverse space
which has to be the same for both the target and the projectile
partons in order for the collisions to occur. The inclusive
cross-section of a double parton scattering can be expressed, as\cite{Gal98}:

\begin{eqnarray}
\lefteqn{\sigma_{DS} = \sum_q \int dx_1 dx_2 dx_3 dx_4 \Gamma_A (x_1,x_2;\hat{b})} \nonumber \\
  & & \Gamma_B (x_3,x_4;\hat{b}) \hat{\sigma}^a(x_1,x_3)
        \hat{\sigma}^b(x_2,x_4) d^2b 
\end{eqnarray}

Where, \( \hat{\sigma}^a(x_1,x_3) \), \( \hat{\sigma}^b(x_2,x_4) \)
are two distinguishable partonic cross sections and \( \Gamma_A
(x_1,x_2;\hat{b}) \) represents the distribution function with
fractional momenta \((x_1,x_2)\) and transverse relative distance with
in the hadron $\hat{b}$.  Double parton distribution functions,
contain in principle detailed information on the hadronic structure
and probes correlations between partons in the hadron in the
transverse structure. These correlations can be assumed to be
negligible if a scattering event is characterized by high
center-of-mass energy. The two-body parton distribution functions
factorize as \( \Gamma (x_1,x_2;\hat{b})=f(x_1) f(x_2) F(\hat{b}) \),
where $f(x)$ is the parton distribution function and \( F(\hat{b}) \)
describes the distribution of parton in the transverse plane. With
these assumptions the cross-section for a double parton collision
leads in the case of two distinguishable parton interactions to the
simplest factorized expression as:

\begin{equation}
\sigma_{DS} = \frac{\sigma^a \sigma^b}{\sigma_{eff}}
\end{equation}

Here \(\sigma^a \) represents the single scattering cross section

\begin{equation}
\sigma^a = \sum_{ij} \int dx_1 dx_2 f_i(x_1) f_j(x_2) \hat{\sigma}_{ij
\rightarrow a}
\end{equation}

where \(f_i(x_1)\) is the standard parton distribution function and
\( \hat{\sigma}_{ij \rightarrow a}\) represents the sub-process cross
section.

The parameter \(\sigma_{eff}=1/\int d^2 \hat{b} F(\hat{b})\), is the
effective cross section and it enters as a simple proportionality
factor in the integrated inclusive cross section for a double parton
scattering \(\sigma_{DS}\).

The experimental value measured by CDF yields \( \sigma_{eff}=1.45 \pm
1.7^{+1.7}_{-2.3}\) mb~\cite{Abe97}. It is believed that it is largely
independent of the center-of-mass energy of the collision and of the
nature of the partonic interactions.

%%%%%%%%%%%%%%%%%%%%%%%%   S e c t i o n   3   %%%%%%%%%%%%%%%%%%%%%%%%

\section{CROSS-SECTION RESULTS}

For the production of Higgs boson at hadron colliders, we consider the
simplest spontaneous broken gauge boson involving exactly one physical
Higgs boson. The elementary cross section for the process \(q\bar{q}
\rightarrow V^* \rightarrow W^\pm\, (or\, Z) + H\) is convoluted with
the appropriate quark and anti-quark distribution functions and
integrated over all appropriate final state variables~\cite{Ell96}.

\begin{eqnarray}
\lefteqn{\sigma(pp \rightarrow V^* \rightarrow W^\pm\, or\, Z + H + X) =} \nonumber \\
 & & \sum_{ij} \int dx_1 dx_2 f_i(x_1) f_j(x_2) \hat{\sigma}(q\bar{q}\rightarrow V^*)
\end{eqnarray}

The sub process cross sections are obtained:

\[
\hat{\sigma}(q\bar{q} \rightarrow WH) = \frac{(G_FM^2_W)^2}{9\pi}
\left|V_{qq'}\right|^2 \frac{p_W}{\sqrt{\hat{s}}}
\frac{3M^2_W+p^2_W}{(\hat{s}-M_W)^2},
\]

\[
\hat{\sigma}(q\bar{q} \rightarrow ZH) = \frac{(G_FM^2_Z)^2}{9\pi}
\left|V^2_q+A^2_q\right|^2 \frac{p_Z}{\sqrt{\hat{s}}}
\frac{3M^2_Z+p^2_Z}{(\hat{s}-M_Z)^2},
\]

Where \(p^2_V=\frac{1}{4\hat{s}} (\hat{s}^2 + M^4_V + M^4_H -
2\hat{s}M^2_V-2\hat{s}M^2_H-2M^2VM^2_H) \) for $V=W,Z$.

\begin{figure}[htb]
\centerline{\includegraphics[width=72mm]{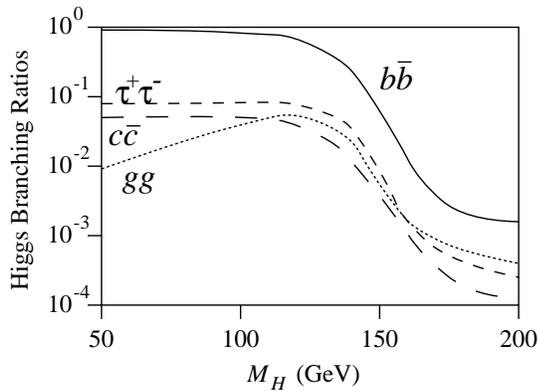}}
\vspace{-24pt}
\caption{Branching ratios of the SM Higgs boson in the mass range for the decay modes $b\bar{b},\, \tau^+\tau^-,\, c\bar{c},\, gg.$}
\label{fig:fig_1}
\end{figure}

In order to devise search strategies for the Higgs boson, it is
important to know the dominant decay channels for different Higgs
boson masses. If the Higgs mass is below the $WW$ threshold then
\(H\rightarrow b\bar{b}\) decay is dominant, as shown in
Fig~\ref{fig:fig_1}.

%\vspace{-24pt}

\begin{figure}[htb]
\centerline{\includegraphics[width=72mm]{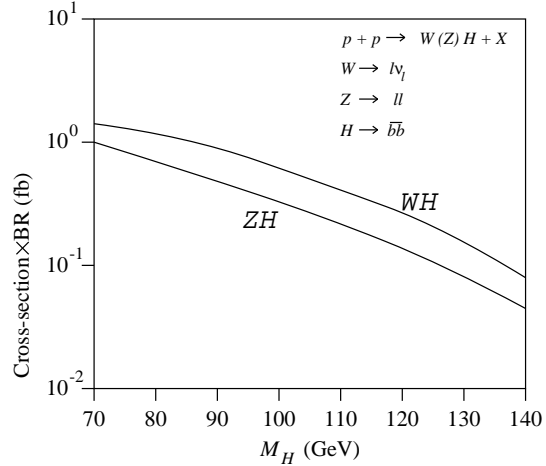}}
\vspace{-24pt}
\caption{Total cross section for $WH$ and $ZH$ production at the LHC
 times the branching ratios of the decay modes, as function of Higgs
 mass.}
\label{fig:fig_2}
\end{figure}

The total cross section for the processes \(p+p \rightarrow WH + X\)
and \(p+p \rightarrow ZH + X\) times the branching ratios of the decay
mode \(H \rightarrow b\bar{b}\), \(W \rightarrow l\nu_l\), \(Z
\rightarrow l^+l^-\) (\(l=e,\mu\)), as a function of Higgs mass at the
LHC energy, is plotted in Fig.~\ref{fig:fig_2}. In our calculation the
integration was performed by VEGAS with MRST parton
distributions~\cite{Mrt03}.

The cross sections for Higgs production in association with $W$ and
$Z$ bosons are not large, but may be useful if high luminosity is
available, since the Higgs can be ``tagged'' by triggering on the weak
boson.

%\vspace{-24pt}

A background to the process \(p+p\rightarrow W^\pm (or\, Z)\, +X\)
with \(W \rightarrow l\nu_l\) and \(H \rightarrow b\bar{b}\), is
represented by the double scattering interactions.  The corresponding
integrated rate is evaluated by combining the expected cross section
for $W$ and $b\bar{b}$ production at LHC energy. Using equation (2)
one obtain the cross section for a double parton collision production
background.

\begin{table}[htb]
%\vspace{-24pt}
\caption{The total cross section for $W^\pm\, (or\, Z)H$ production at
         the LHC times the branching ratios of the decay modes
         compared with the double scattering parton background.
         $\sigma_{SS}$ is cross Section for a Single Scattering
         process and $\sigma_{DS}$ is the cross section for double
         parton scattering background}
\begin{center}
\begin{tabular}{ccc} \hline
  Higgs Production & $\sigma_{SS}\times BR$ & $\sigma_{DS}$ \\ \hline
  \(q\bar{q}\rightarrow WH\) & 0.5 pb & 1.4 nb \\
  \(q\bar{q}\rightarrow ZH\) & 0.3 pb & 0.17 nb \\ \hline

\end{tabular}
\end{center}
\label{tbl:tbl_1}
\end{table}
%\vspace{-24pt}

It is clear that one obtains a cross section in double scattering
background three orders of magnitude larger that the expected signals
from Higgs decay.

The large rate of $b\bar{b}$ pairs expected at the LHC gives
rise to a relatively large production of pair in the process
underlying the $W$ production
 as a very promising channel to detect
the production of the Higgs boson.

\begin{figure}[htb]
\centerline{\includegraphics[width=72mm]{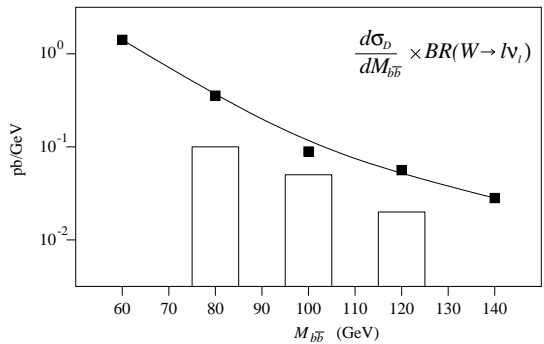}}
\vspace{-24pt}
\caption{Double parton scattering background to the Higgs boson
         production as a function of the $b\bar{b}$ invariant mass
         compared to the expected Higgs signal for three possible
         values of the Higgs mass 80, 100 and 120 GeV.}
\label{fig:fig_3}
\end{figure}

Figure \ref{fig:fig_3} shows the double parton scattering background
to the Higgs boson production as a function of the bb invariant mass
compared to the expected signal for three possible values of the Higgs
mass. The double parton scattering process remains a rather
substantial component of the background to Higgs production.

%\vspace{-24pt}

%%%%%%%%%%%%%%%%%%%%%%%   D I S C U S S I O N   %%%%%%%%%%%%%%%%%%%%%%%

\section{SUMMARY}

In this work we have computed the most promising channel to detect the
production of the SM Higgs particle at the LHC with a mass below about
135 GeV, where the Higgs decays into $b\bar{b}$ final states is dominant.

The production of the Higgs boson, in the intermediate mass range,
namely the final state with a $b\bar{b}$ pair and with isolated
lepton, is affected by a sizable background due to double parton
collision process.

Although the double parton scattering background cross section is a
decreasing function of the invariant masses of the $b\bar{b}$ pair,
the relatively large value of the invariant mass required to the
$b\bar{b}$ pair to be assigned to the Higgs decay is not large enough,
at LHC energies, to allow one to neglect the double parton scattering
background.

However, in the hadron collider environment, the large QCD backgrounds
may cause the observation impossible for those events if one or more
of the gauge bosons decay hadronically. Individual channels with
hadronic decays should be studied on case by case.

It is useful to point some of the inherent uncertainties which affect
the final results. The most significant are: (i) the lack of precise
knowledge of the parton distribution at small $x$, which is important
for the intermediate mass Higgs and (ii) the effect of unknown higher
order perturbation QCD corrections.

%%%%%%%%%%%%%%%%%   T H E   B I B L I O G R A P H Y   %%%%%%%%%%%%%%%%%

%%%%%%%%%%%%%%%%%%%%%%%%%%%%%%%%%%%%%%%%%%%%%%%%%%%%%%%%%%%%%%%%%%%%%%
%%%%%%%%%%%%%%%%%%%%%%%%%%%%%%%%%%%%%%%%%%%%%%%%%%%%%%%%%%%%%%%%%%%%%%
%%%%%%%%%%%%%%%%%%%%%%%%%%%%%%%%%%%%%%%%%%%%%%%%%%%%%%%%%%%%%%%%%%%%%%


\begin{thebibliography}{19}
% [1]
\bibitem{Hgg64} P. W. Higgs, Phys. Lett. 12(1964) 13 2;  Phys. Rev. Lett. 13
                (1964) 508; Phys. Rev. 145(1966) 1156.

%[2]
\bibitem{Dai95} J. Dai, J. F. Gunian and R. Vega, Phys. Rev. Lett.  71 (1993)
                2699; A. stange, W. Marciano and S. S. D. Willenbrock, Phys.
                Rev. D49 (1994) 1354; J. F. Ganion and T. Han , Phys. Rev. D51
                (1995) 1051.

%[3]
\bibitem{Lnd82} P. V. Landshoff and J. C. Polkinghorne, Phys. Rev. D18 (1978)
                3344; C. N. Paver and  D. Treleani, Nuovo Cimento A70 (1982)
                215; F. Halzen , P. Hoyer and W. J. Stirling Phys. Lett., 188B 
                (1987) 375; F. Abe et. al. (CDF Collaboration), Phys. Rev.
                Lett.  79 (1997) 584; Phys. Rev. D56 (1997) 3811.

%[4]
\bibitem{Gal98} G. Galucci and D. Treieani, Phys. Rev. D57 (1998) 503.

%[5]
\bibitem{Abe97} F. Abe et al. (CDF Collaboration), Phys. Rev. Lett. 79 (1997)
                584; Phys. Rev. D56 (1997) 3811.

%[6]
\bibitem{Ell96} R. K. Ellis, W. J. Stirling and B. R. Webber, QCD and Collider Physics,  Cambridge University Press, Cambridge (1996).

% [7]
\bibitem{Mrt03} A. D. Martin, R. G. Roberts,  W. J. Stirling and R. S. Thorne,
                Eur. Phys. J. C28 (2003) 455 [hep-ph/0211080].

\end{thebibliography}
\end{document}